\documentclass[twocolumn,showpacs,preprintnumbers,amsmath,amssymb,prl]{revtex4}
\usepackage{graphicx}
\usepackage{bm}

\usepackage[usenames]{color}

\newcommand{ \be}{\begin{equation}}
\newcommand{ \ee}{\end{equation}}
\newcommand{\beq}{\begin{eqnarray}}
\newcommand{\eeq}{\end{eqnarray}}
\newcommand{\bem}{\begin{pmatrix}}
\newcommand{\eem}{\end{pmatrix}}
\newcommand{\bmx}{\begin{array}}
\newcommand{\emx}{\end{array}}

\begin{document}

\title{How to quantify structural anomalies in fluids ?}

\author{Yu. D. Fomin}
\affiliation{Institute for High Pressure Physics, Russian Academy
of Sciences, Troitsk 142190, Moscow Region, Russia}

\author{B. A. Klumov}
\affiliation{High Temperature Institute, Russian Academy of
Sciences, 125412, Izhorskaya 13/2, Russia}

\author{V. N. Ryzhov}
\affiliation{Institute for High Pressure Physics, Russian Academy
of Sciences, Troitsk 142190, Moscow Region, Russia}
\affiliation{Moscow Institute of Physics and Technology, 141700
Moscow, Russia}

\author{E. N. Tsiok}
\affiliation{Institute for High Pressure Physics, Russian Academy
of Sciences, Troitsk 142190, Moscow Region, Russia}

\date{\today}

\begin{abstract}
Some fluids are known to behave anomalously. The so-called
structural anomaly which means that the fluid becomes less
structures under isothermal compression is among the most
frequently discussed ones. Several methods for quantifying the
degree of structural order are described in the literature and are
used for calculating the regions of structural anomalies. It is
implied that all of  the structural order determinations yield
qualitatively identical results. However, no explicit comparison
was made. This paper presents such a comparison for the first
time. the results of some definitions are shown to contradict the
intuitive notion of a fluid. On the basis of this comparison we
show that the structural anomaly can be most reliably determined
from the behavior of the excess entropy.
\end{abstract}

 \maketitle

\section{I. Introduction}

It is well known that some liquids demonstrate anomalous
properties \cite{book}. The most common example of an anomalous
liquid is water \cite{wateranomalies}. Water has a region of
density - temperature parameters where the diffusion coefficient
increases with density (diffusion anomaly), thermal expansion
coefficient has a negative sign (density anomaly) and the liquid
becomes less structured with increasing density (structural
anomaly). Another common example of an anomalous liquid is silica
which also has regions of anomalous diffusion, density and
structure \cite{deben}. Having analyzed a wide set of data,
Errington and Debenedetti came to the conclusion that regions of
different anomalies form nested domains \cite{orderanom}. In
particular, it was shown that in the case of water, the order of
anomalies is as follows: the region of anomalous density is inside
the diffusion anomaly region and both of them are inside of the
region of the structural anomaly. In the case of silica, the order
of the anomalies is different: the density anomaly is inside the
structural anomaly and both of them are inside the diffusion
anomaly \cite{deben}. Later on it was shown that the order of the
density anomaly and the structural anomaly determined by the
anomaly of excess entropy is strictly defined by the thermodynamic
relations \cite{str-den}, while there is no such relation between
the diffusion anomaly region and the other anomalies
\cite{silicalike}. So, the diffusion anomaly can take any place in
the $\rho - T$ plane with respect to other anomalies
\cite{silicalike,jcpsequence,specialtopics}.

Among the systems which demonstrate an anomalous behavior, the
simplest ones to study are the so called core-softened systems
(see, for example,
\cite{hemmer1,hemmer2,netz,buldyrevstanley,jagl1,jagla,jagl2,jagl3,
jagla1,jagla2,silicalike,jcpsequence,specialtopics,we1,we2,we4,we3,RCR,barboska1,barboska,
barb1,barb2,barb3,buld1,buld2,buld3,buld2009,buld2011,franzese1,franzese2,fr3,fr4,nature,ind}).
Many core-softened systems demonstrate an anomalous behavior due
to the presence of two length scales in the interaction potential.
As a result, two locally preferred structures are possible: low
and high density ones. The competition between these structures
leads to the appearance of anomalous behavior. Later on it was
shown that even some models with one scale can behave anomalously
\cite{prestipino}. Here, anomalous behavior is related to the
shape of the force as a function of distance.

It is widely believed that in the core-softened systems the
hierarchy of anomalies is of the water-like type. However,
recently it has been shown that the order of the region of
anomalous diffusion and the regions of density and structural
anomalies may be inverted depending on the parameters of the
potential and may have the silica-like or some other sequences
\cite{silicalike,jcpsequence,specialtopics}. Several definitions
of structural anomaly are available in the literature. Some of the
definitions directly refer to the structural properties. They rely
on different structure-dependent parameters which are usually
called the order parameters (do not confuse with the order
parameters in the theory of phase transitions)
\cite{orderanom,barboska1,barboska}. One expects that in ordinary
systems the order parameter increases with density along the
isotherm. The region of densities, where the order parameter
decreases under densification, will be an anomalous region.

Other definitions of the structural anomaly relate it to the
excess entropy $S_{\rm ex}=S - S_{\rm id}$ where $S$ is the total
entropy and $S_{\rm id}$ is the entropy of ideal gas at the same
density and temperature. Entropy defines the number of states
accessible to the system. So, the less ordered the system, the
larger the excess entropy. In a typical liquid the excess entropy
decreases with isothermal increasing of density because it becomes
more ordered. In an anomalous liquid there is a region where
$S_{\rm ex}$ increases with density.

The appearance of structural anomaly in liquids was widely
discussed in the literature (see, for example,
\cite{silicalike,jcpsequence,specialtopics,we1,we2,we4,we3,RCR,barboska1,barboska,
buld1,buld2,buld2009,franzese1,franzese2}). However, different
authors use different definitions for the structural anomaly. It
was assumed that all of the definitions give qualitatively similar
results. However, we are not aware of any verification of this
assumption. The main goal of this paper is to find the most
consistent way to determine the region of structural anomaly. In
order to do this we calculate the structurally anomalous regions
by different definitions and compare them to each other. We
believe that this study will clarify the relation between
different definitions of structural anomaly.

\section{II. System and Methods}

We study a core-softened potential system introduced in our
previous publications
\cite{silicalike,jcpsequence,specialtopics,we1,we2,we4,we3,RCR},
namely the repulsive shoulder system (RSS) which is defined by the
following interparticle potential:
\begin{equation}
  \Phi= \varepsilon (\sigma/r)^{14}+ \varepsilon \cdot \left [
  1-\tanh(k \cdot (r- \sigma_1))\right ]/2, \label{1}
\end{equation}
where $k=10$, $\sigma_1=1.35$. Thereafter of this paper, we use
the dimensionless quantities: $\tilde{{\bf r}}\equiv {\bf r}/
\sigma$, $\tilde{P}\equiv P \sigma^{3}/\varepsilon ,$
$\tilde{V}\equiv V/N \sigma^{3}\equiv 1/\tilde{\rho}$. As only
these reduced variables are used, we will omit the tildes.

In our previous works, we presented the phase diagram of this
system \cite{we1,we2,we4,we3} and discussed the anomalous
behavior. The order of anomalies corresponding to the case of the
water - density anomaly region is inside the diffusion anomaly one
and both of them are inside the structural anomaly. In the later
works we investigated the influence of the shape of the potential
on anomalies in the system. It has been shown that if the width of
the repulsive shoulder increases (i.e. increases $\sigma_1$), the
anomalies start to disappear. At $\sigma_1=1.55$, the diffusion
anomaly vanishes. If $\sigma_1=1.8$, the density anomaly also
vanished and only the structural anomaly preserves
\cite{jcpsequence,we2}.

While increase of the repulsion depresses anomalies, the
attractive forces stabilize them \cite{silicalike,specialtopics}.
If the attractive well is added to the potential (\ref{1}), the
anomalous regions will extend to higher temperatures. Moreover,
the sequence of regions changes, and at some attractive well depth
the order of anomalies becomes silica-like
\cite{silicalike,specialtopics}.

In the present work we simulated a system of $4000$ particles in a
cubic box in molecular dynamics. The system was simulated in the
canonical ensemble - constant number of particles $N$, volume $V$
and temperature $T$. The timestep was set at $0.005$ reduced units
of time. The system was equilibrated for $3 \cdot 10^6$ steps. The
production run was set at $1 \cdot 10^6$ steps. Each $500$ steps
the current configuration was stored in order to compute the
functions of interest. The density varied from $\rho=0.15$ up to
$0.75$ with the step $d \rho =0.05$ and temperature from $T=0.1$
up to $0.7$.

All simulations in this work were done using lammps simulation
package \cite{lammps}.

In our previous works we defined the structural anomaly via the
anomaly of  the excess entropy $S_{\rm ex}$. Here, since we wish
to compare different definitions of the structural anomaly
regions, we add some more criteria.

We compute the structural anomaly via the pair entropy \be s_2=-2
\pi \rho \int{(g(r) \cdot \ln(g(r))-g(r)+1}) \cdot r^2 dr,
\label{2} \ee The pair entropy can be used not only for
characterizing anomalous properties of a liquid \cite{s2gauss} but
also as the "one-phase" criteron of freezing
\cite{s2freeze,s2freeze1}.

The translational order can be measured with the use of the order
parameter $\tau$ introduced by Truskett et al. \cite{deben1} and
modified for water by Errington and Debenedetti \cite{orderanom}:
\be
  \tau = \frac{1}{\xi_c} \int_0 ^{\xi_c} |g(\xi)-1| d \xi, \label{3}
\ee
where $\xi=r \cdot \rho^{1/3}$ is the interparticle separation
scaled by the mean interparticle distance, $g(\xi)$ is the radial
distribution function and $\xi_c$ is a scaled cutoff distance. In
this work, we use $\xi_c=\rho^{1/3}L/2$, where $L=V^{1/3}$.

Other parameters considered here are associated with the
orientational local order. To define the local orientational
properties of the repulsive shoulder system we use the bond order
parameter method, which has been widely used in the context of
condensed matter
physics~\cite{stein,accuratecryst,orderglass,nucllj,complexpart,gasser1,torqa},
hard sphere systems~\cite{hs_init,troadec,aste,hskl,auer}, complex
(dusty) plasmas~\cite{3Da,mitic,3Db,pu,khr}, colloidal
suspensions~\cite{coll0,coll1,coll2}, etc.

According to this method the rotational invariants (RIs) of the
rank $l$ of both the second $q_l(i)$ and the third $w_l(i)$ orders
are calculated for each particle $i$ in the system from the
vectors (bonds) connecting its center with the centers of its
$N_{\rm nn}(i)$ nearest neighboring particles: \be q_l(i) = \left
( {4 \pi \over (2l+1)} \sum_{m=-l}^{m=l}
\vert~q_{lm}(i)\vert^{2}\right )^{1/2} \label{4} \ee \be
w_l(i) = \hspace{-0.8cm} \sum\limits_{\bmx {cc} _{m_1,m_2,m_3} \\_{ m_1+m_2+m_3=0} \emx} \hspace{-0.8cm} \left [ \bmx {ccc} l&l&l \\
m_1&m_2&m_3 \emx \right] q_{lm_1}(i) q_{lm_2}(i) q_{lm_3}(i),
\label{wig} \ee \noindent where $q_{lm}(i) = N_{\rm nn}(i)^{-1}
\sum_{j=1}^{N_{\rm nn}(i)} Y_{lm}({\bf r}_{ij} )$, $Y_{lm}$ are
the spherical harmonics and ${\bf r}_{ij} = {\bf r}_i - {\bf r}_j$
are the vectors connecting the centers of particle $i$ and $j$.

In Eq.(\ref{wig}), $\left [ \bmx {ccc} l&l&l \\ m_1&m_2&m_3 \emx
\right ]$ are the Wigner 3$j$-symbols, and the summation in the
latter expression is performed over all of the indexes $m_i
=-l,...,l$ satisfying the condition $m_1+m_2+m_3=0$. Here, for
detecting close packed crystalline structures, we calculate the
rotational invariants $q_4$, $q_6$, $w_4$ and $w_6$ for each
particle using the fixed number of nearest neighbors  $N_{\rm
nn}=12$. Particles whose coordinates in the 4-dimensional space
$(q_4,q_6,w_4,w_6)$ are sufficiently close to those of the ideal
face-centered cubic (fcc), hexagonal close packed (hcp),
icosahedral (ico), etc lattice are counted as fcc-like (hcp-like,
ico-like) particles. By calculating the bond order parameters, it
is easy to  identify the disordered (liquid-like) phase (for
instance, such particles have the mean bond order parameter
$q_6^{\rm liq} \simeq N_{\rm nn}^{-1/2} \simeq 0.29 \ll q_6^{\rm
fcc/hcp/ico}$, where $N_{\rm nn}=12$). By varying the number of
nearest neighbors $N_{\rm nn}$ and the rank $l$ of bond order
parameter, it is possible to identify the lattice type,
quasicrystalline order, distorted hcp/fcc/ico modifications,
liquid-like particles, etc. The values of $q_l$ and $w_l$ for some
crystals are presented in Table~\ref{t1}. Comparing the calculated
values of the corresponding $q_l$ and $w_l$ with their ideal
values we can estimate the local order in the liquid.

\begin{table}[!ht]
  \centering
  \caption{Rotational invariants $q_l$ and $w_l$ ($l=4,~6$) of a few
  perfect crystals calculated via the fixed number of nearest neighbors (NN): hexagonal
  close-packed (hcp), face centered cubic (fcc), face centered tetragonal (fct),
  icosahedron (ico), body-centered cubic (bcc), simple hexagonal (sh),
  simple cubic (sc) and diamond (dia).}
\begin{tabular}{|c|c|c|c|c|}
\hline 
c rystalline structure & \quad $q_{4}$ & \quad $q_{6}$ & \quad
$w_{4}$ & \quad $ w_{6}$ \\ \hline hcp (12 NN) & 0.097 & 0.485 &
0.134  & -0.012 \\ \hline fcc  (12 NN) & 0.19  & 0.575  & -0.159
&  -0.013 \\ \hline fct  (12 NN) & 0.225  & 0.51  &  0.11  &
0.018 \\ \hline ico (12 NN) & $1.4 \times 10^{-4}$ & 0.663 &
-0.159  & -0.169 \\ \hline bcc ( 8 NN) & 0.5 & 0.628 & -0.159   &
0.013 \\  \hline bcc (14 NN) & 0.036 & 0.51 & 0.159   & 0.013 \\
\hline sh   ( 8 NN) & 0.53 & 0.5 & 0.134   & 0.0475 \\  \hline sc
(  6 NN) & 0.76 & 0.35 & 0.159   & 0.013 \\ \hline
dia  ( 4 NN) & 0.51 & 0.628 & -0.159   & 0.013 \\
\hline 
\end{tabular}
\label{t1}
\end{table}

To quantify the global orientational order, it is convenient to
use the metrics associated with the cumulative distributions of
the normalized probability distribution functions (PDF) $P(q_l)$
and $P(w_l)$ for different order parameters $q_l$ and $w_l$
\cite{pu,hskl,lj1,lj2}. For instance, the cumulative function
$C_q^l$ of  $P(q_l)$ is defined from: \be C_q^l (x) \equiv
\int_{-\infty}^x P(q_l)dq_l \ee Evidently, $C_q^l (x)$ is the
abundance of particles, having $q_l < x$ and $C_q^l (\infty)
\equiv 1$. The relevant order parameter $Q_l^c$ is the position of
the half-height of the cumulative distribution $C_q^l$, so that
$C_q^l (Q_l^c) \equiv 1/2$. Here, we use the cumulants $Q_6^c$ and
$W_6^c$ associated with the particle distributions over bond order
parameters $q_6$ and $w_6$, respectively as indicators of the
structural anomaly.

Having defined the order parameters one can determine the region
of structural anomaly with respect to these parameters. It is
intuitively clear from the term "order parameter" that the larger
value of the order parameter corresponds to the larger structural
order. Therefore, in normal fluids the order parameters should
increase with increasing density along the isotherm, and if one
finds a region of the isotherm where the order parameter
decreases, this region will be structurally anomalous.

The opposite situation is observed in the case of excess entropy.
The excess entropy decreases in case of a normal liquid, so the
anomalous region is defined as the region where $S_{\rm ex}$
increases.

\section{III. Results and Discussion}

\begin{figure}
\includegraphics[width=8.4cm]{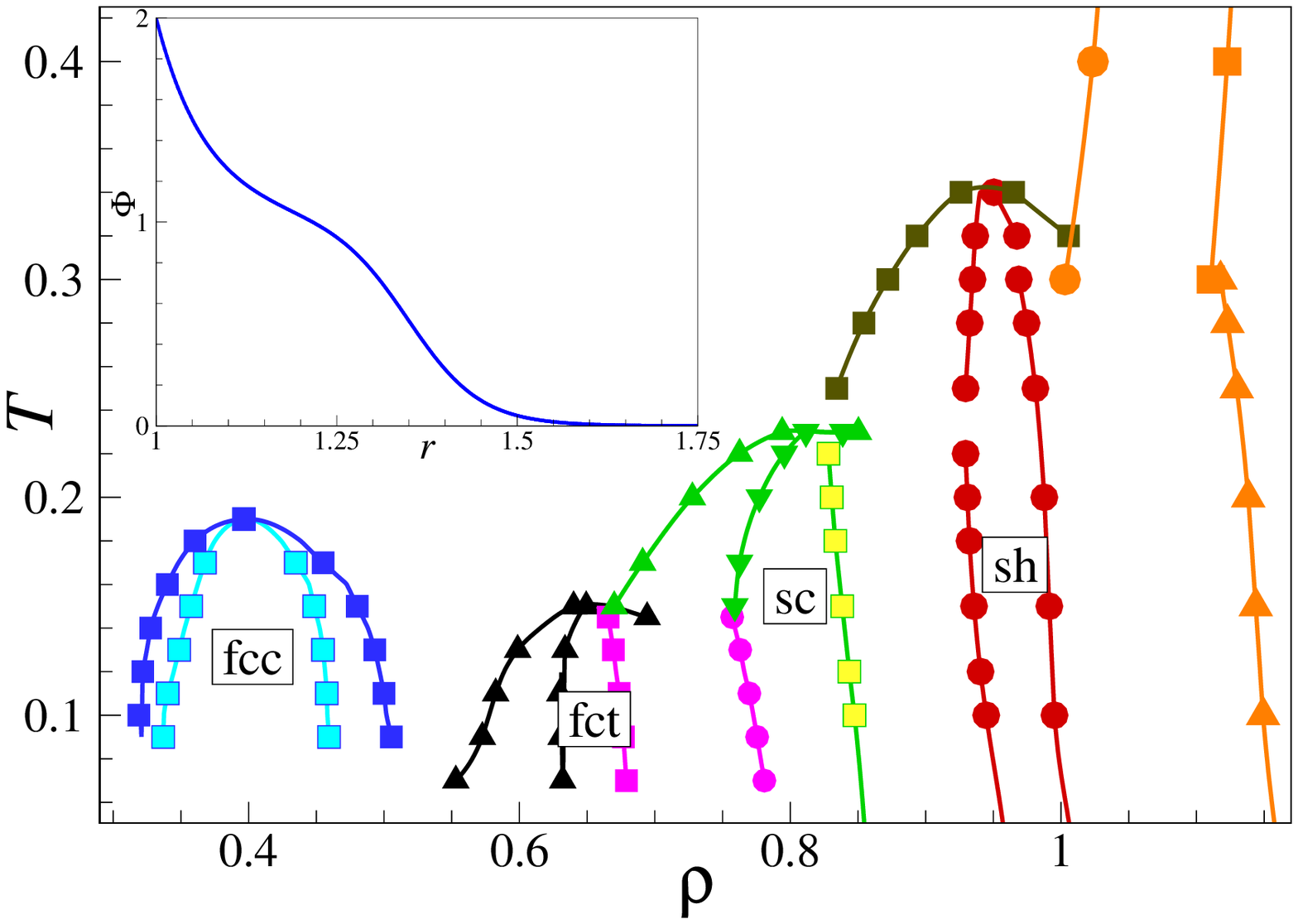}
\includegraphics[width=8.4cm]{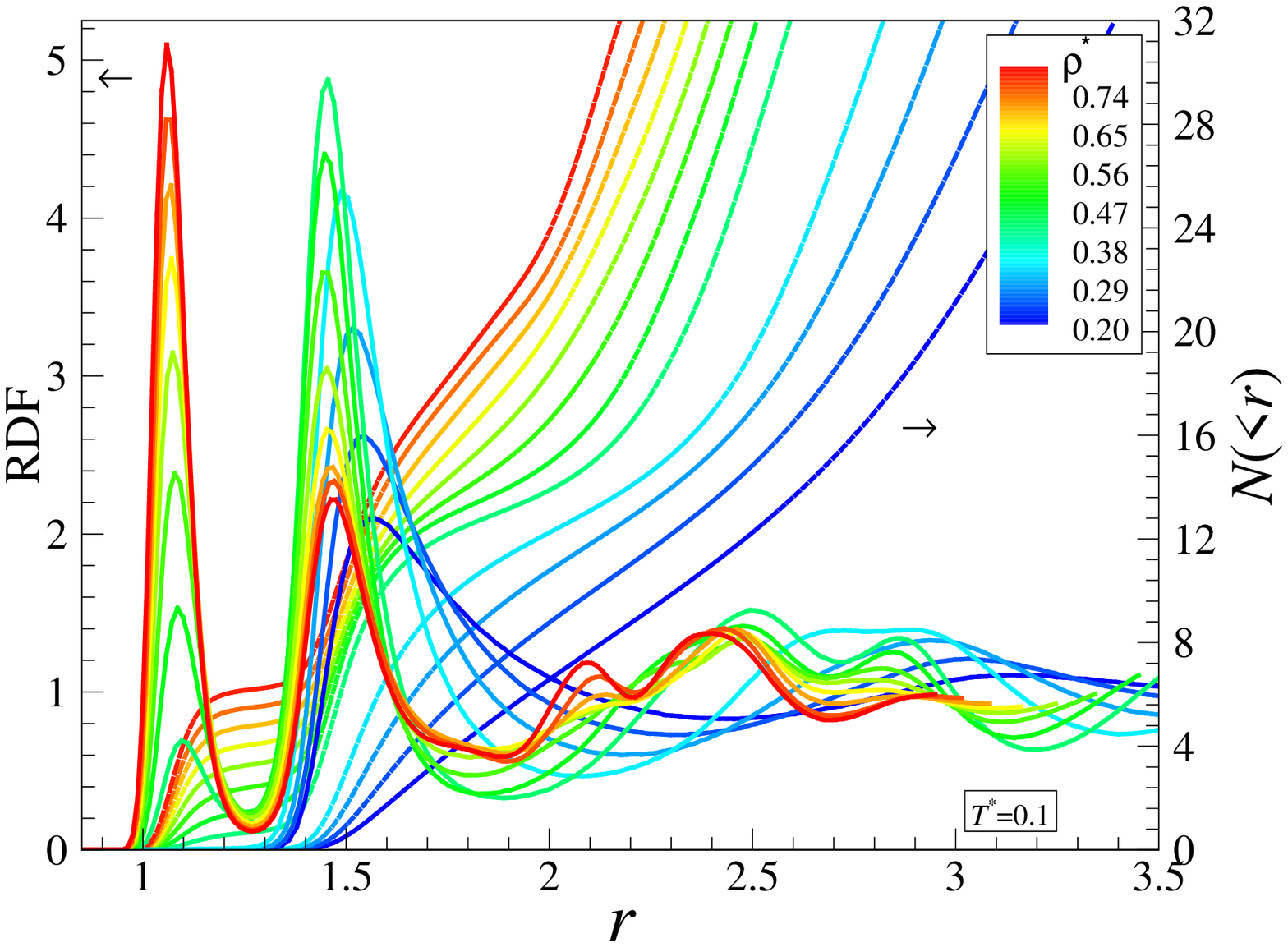}
\caption{\label{fig:fig1} (Color online). Top panel: Phase diagram
of the repulsive shoulder system (RSS) with potential (1). Domains
of different crystalline order (face-centered cubic (fcc),
face-centered tetragonal (fct)), simple cubic (sc) and simple
hexagonal (sh)) are indicated. The inset shows the interaction
potential (Eq. 1). Bottom panel: the radial distribution function
$g(r)$ and its cumulative functions $N(<r)$: ($N(<r) \equiv 4\pi
\rho \int_0^r r'^2 g(r') dr'$ is the mean number of particles
inside the sphere of radius $r$) at different densities $\rho$;
the curves are color-coded depending on the $\rho$ value. The RSS
dimensionless temperature $T = 0.1$.} \label{fig1}
\end{figure}

In order to study the region of structural anomaly defined by
different quantities, first of all, we need to recall the phase
diagram of the system which is shown in Fig.~\ref{fig:fig1}(top
panel). This phase diagram was already published in our previous
papers. We show it here again for the sake of completeness. In the
inset, the radial distribution function $g(r)$ and the number of
particles $N(<r)$ in the spherical volume of radius $r$, at
$T=0.1$ are shown for the set of densities.

First we consider the orientational order parameters $Q_6^c$ and
$W_6^c$. In this study we use the fixed number of nearest
neighbors $N_{\rm nn}=12$ for calculating the bond order
parameters $q_l$ and $w_l$ (and their cumulative measures $Q_6^c$,
$W_6^c$ as well). As is clearly seen from the cumulative
distributions $N(<r)$ plotted in Fig~\ref{fig1}, in the considered
density range $\rho= 0.2 \div 0.75$ the first (at low values $\rho
< 0.5$) and both first and second shells (at $\rho > 0.6$), the
mean number of particles is close to 12. So, the parameters $q_l$
and $w_l$ reflect the orientational ordering of the RSS system;
according to the definition, these parameters include many body
correlations (which depend on $N_{\rm nn}$). It is intuitively
clear that the orientational bond order parameters are much more
sensitive to the local structure rearrangement at the RSS
densification than the metrics which are based on the two-point
correlations properties only ($g(r)$, $\tau$, $s_2$).

\begin{figure}
\includegraphics[width=8.4cm]{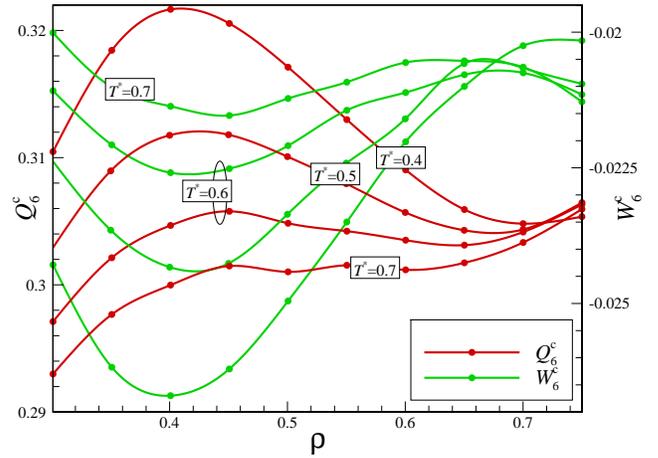}
\caption{\label{fig:fig2} (Color online). Orientational order
parameters associated with the cumulative functions (see text for
details)  of  the PDF of $Q_6^c$ (red lines) and $W_6^c$ (green
lines) calculated by using $N_{\rm nn} =12$ versus RSS
dimensionless density $\rho$ at different temperatures $T$
(indicated on the plot).}
\end{figure}

Fig.~\ref{fig:fig2} shows the parameters $Q_6^c$ and $W_6^c$ for a
set of isotherms. One can see that they behave in opposite ways:
while $Q_6^c$ increases in the normal regime and decreases in
anomalous one, $W_6^c$ normally decreases and increases in the
anomalous region. The points of extrema of these parameters almost
coincide. However, a small difference still takes place. If these
points are placed in the phase diagram, the anomalous region will
extend to the right above the crystal phases. Intuitively, the
anomalies can be explained the influence of the soft core of the
potential. However, at the densities of the minima of $Q_6^c$ or
maxima of $W_6^c$, the influence of the soft core must be
relatively small. So, these orientational parameters do not fit
the intuitive expectations.

In order to elucidate this point, it was proposed to determine the
left-hand branch (low densities) of the region of structural
anomaly is through the orientational order parameter $Q_6^c$ while
the right-hand branch (high densities), by the translational
parameter $\tau$ \cite{orderanom}. However, it looks to be
nonselfconsistent: if we define the structural anomaly as an
anomaly in some parameter it should be expected that the same
parameter gives the whole anomalous region.

\begin{figure}
\includegraphics[width=8.4cm]{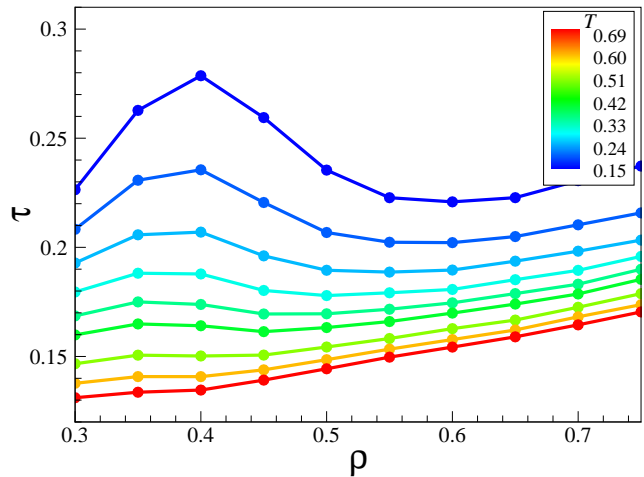}%
\caption{\label{fig:fig3} (Color online). Translational order
parameter $\tau$ along the following isotherms (from bottom to
top): $0.15,0.2,0.25,0.3,0.35,0.4,0.5,0.6,0.7$.}
\end{figure}

The region of structural anomaly can be defined through the minima
and maxima of the translational order parameter $\tau$, as it was
done, for example, in Refs. \cite{barboska1,barboska}.
Fig.~\ref{fig:fig3} shows the translational order parameter $\tau$
along a set of isotherms. One can see that at low enough
temperatures it demonstrates the maximum and minimum; so there is
a region of anomalous behavior of this parameter. If one takes the
points of maxima and minima of $\tau$ one can see that as the
temperature increases the anomalous region goes to the left, i.e.
toward the lower densities. Moreover, the maximum temperature of
this anomaly is rather high - $T_{\rm max}=0.6$. Intuitively one
relates the presence of anomalous behavior to the presence of two
length scales in the potential. In particular, two scales lead to
the formation of the low-density FCC phase in the phase diagram
(Fig.~\ref{fig:fig1}). But the $\tau$ anomaly (as well as the
$q_6$ and $w_6$ anomalies) exists at quite high temperatures when
the two scales do not play so important role. So, this order
parameter also seems to contradict to our intuitive expectations.

\begin{figure}
\includegraphics[width=8.4cm]{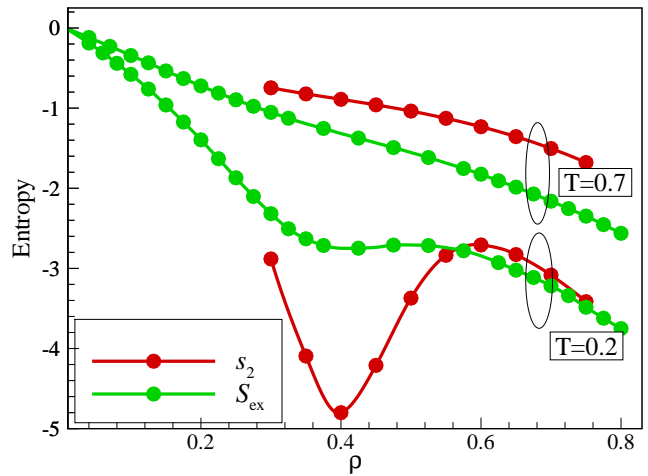}
\caption{\label{fig:fig4} (Color online). Comparison of total the
excess entropy $S_{\rm ex}$ and the pair entropy $s_2$ at $T=0.7$
and $T=0.2$. The ovals unite the curves corresponding to the same
temperature.}
\end{figure}

\begin{figure}
\includegraphics[width=8.4cm] {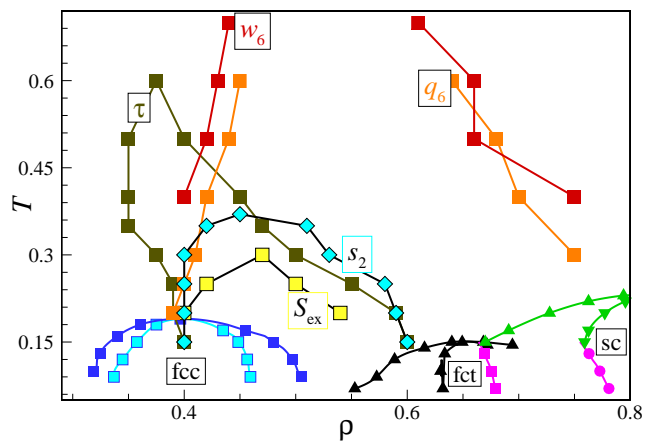}
\caption{\label{fig:fig5} (Color online). The regions of
structural anomalies corresponding to all of the all definitions
described in the text, placed in the phase diagram. The curves of
$q_6$ and $w_6$ are cut at the top since they go into very high
temperatures which are beyond the interest of the present study.}
\end{figure}

Next we consider the definition of the structural anomaly by the
excess entropy $S_{\rm ex}$ and the pair contribution to the
excess entropy $s_2$. It is commonly supposed that the pair
entropy is a good approximation to the excess entropy. Usually it
is referred to the Lennard-Jones system \cite{baranyai} where the
pair entropy accounts for approximately $80-85$ percent of the
total excess entropy. However, as it was shown in our previous
publication, in the case of core-softened systems $s_2$ strongly
deviates from the total excess entropy \cite{we4}.
Fig.~\ref{fig:fig4} compares the total excess entropy with the
pair entropy at a relatively high temperature ($T=0.7$ which is
higher then the anomalous region) and at a low temperature
($T=0.2$). One can see that the difference between the two curves
is rather large. For $T=0.7$ it reaches $40 \%$. At low
temperatures the largest difference is approximately $75 \%$ at
$\rho=0.4$. At the same time, at high densities under $T=0.2$ the
pair entropy very closely follows the $S_{\rm ex}$ curve. One can
conclude that $s_2$ can be close to the total excess entropy but
its real location with respect to $S_{\rm ex}$ is rather
unpredictable. So, it is not a good quantity to refer to. However,
since $s_2$ is actively studied in the literature we also computed
it and found the structural anomaly region related to $s_2$. The
corresponding curve is shown in Fig.~\ref{fig:fig5} where all
definitions of the structural anomaly are shown for comparison.

Typically radial distribution function is used for characterizing
the structure of liquid. The translational order $\tau$ and the
pair excess entropy $s_2$ represent two different ways to quantify
this structure. However, one can see that these two ways give very
different results which clearly indicates that a solid theoretical
basis is needed for quantifying the liquid structure by radial
distribution function derivatives.

Finally we determined the structural anomaly through the maxima
and minima of $S_{\rm ex}$ as it was done in our previous
publications
\cite{silicalike,jcpsequence,specialtopics,we1,we2,we4,we3}. The
resulting curve is shown in Fig.~\ref{fig:fig5}.

It is seen from Fig.~\ref{fig:fig5} that all regions differ
widely. This leads to an obvious conclusion that if someone meets
"a structural anomaly" in the literature one should mention with
respect to which quantity this anomaly is defined. However, to
date this question was not clarified. As a result the confusion is
possible. For example, it is common in the literature to define
the "water like order of anomalies" (density anomaly is inside the
diffusion anomaly and both of them are inside the structural
anomaly) and the "silica-like order" (density anomaly is inside
the structural anomaly and both of them are inside the diffusion
anomaly). However, if one considers, for example, Fig. 11 (d) of
Ref. \cite{induska} where the anomalous regions of SPC/E water are
shown one can see that the region of anomalous $s_2$ is located
inside the anomalous diffusion region. Another region of the
structural anomaly is defined in Ref. \cite{induska} by the local
tetrahedral order parameter $q_{\rm tet}$ associated with the atom
$i$ is defined as
\begin{equation}
q_{\rm tet} = 1 - \frac{3}{8}\sum_{j=1}^{3}\sum_{k=j+1}^{4}(\cos
\psi_{jk}+1/3)^2,
\end{equation}
where $\psi_{jk}$ is the angle between the bond vectors ${\bf
r}_{ij}$ and ${\bf r}_{ik}$, where $j$ and $k$ label the four
nearest neighbor atoms of the same type\cite{str-den}. $q_{\rm
tet}$ characterizes the orientational order and in this sense is
close to $Q_6^c$. This region covers the diffusion anomaly. It
means that depending on the definition of the structural anomaly,
one can observe in SPC/E water or water-like or silica-like
behavior. Thinking about this problem one finds very unusual to
observe the silica-like behavior of water. Apparently, it is
necessary to use unambiguous definitions.

In order to decide which definition of the structural anomaly is
the most adequate, we propose to use the following reasoning.

As it was shown in Ref. \cite{str-den} (see also
\cite{silicalike,jcpsequence,specialtopics}), there are some
thermodynamic relations between the regions of anomalous density
and the excess entropy (note, that in Ref. \cite{str-den} similar
relation was erroneously established for the diffusion anomaly
region \cite{we4}.) If the derivative of $S_{\rm ex}$ with respect
to the logarithm of density is considered
\begin{equation}
  \frac{\partial S_{\rm ex}}{\partial \ln (\rho)}>c, \label{der}
\end{equation}
$c=0$ will correspond to the structural anomaly defined through
excess entropy, while $c=1$ is the condition for the density
anomaly \cite{str-den}. It means that the region of anomalous
density is always inside the region of anomalous excess entropy.
So, the thermodynamically consistent relation between the regions
of two anomalies is established. Note, that in the considered set
of anomalies (density, structure and diffusion), two of them are
defined via thermodynamic properties (density and structure via
$S_{\rm ex}$), and diffusion has the dynamic nature. The
alternative definitions via the order parameters are not
thermodynamically consistent and have the structural nature. As a
result, they do not obey any relations like Eq. (\ref{der}) which
makes them less convenient to analyze the relation between the
anomalies. It is worth to emphasize that the derivation of Eq.
(\ref{der}) is purely thermodynamical and does not contain any
reference to a concrete potential. This relation between the
excess entropy anomaly and the density anomaly is based on the
exact thermodynamic relations, so it is valid for any system.
Similar relation for the diffusion anomaly which was proposed in
Ref. \cite{str-den} was based on the Rosenfeld scaling relations
between the diffusion coefficient and the excess entropy
\cite{ros1,ros2}. However, as it was shown later \cite{we4}, the
Rosenfeld relation breaks down in anomalous regions and,
therefore, it cannot be used for identifying the relative location
of the regions of density, structure and diffusion anomalies.

Although the orientational order parameters $Q_6^c$ and $W_6^c$
are the most powerful tool for quantifying the structure, there
are some difficulties in applying these metrics to fluids. The
most important problem is a correct definition of the nearest
neighbors in liquid. Although the determination of the nearest
neighbor of a particle is of great importance there is no unique
definition in the literature. Several definitions are available in
the literature \cite{nucllj,okabe,pnas,daan} and the application
of different definitions can alter the results, especially in the
region of anomalies where the presence of two scales in the
potential is most important. Although the number of nearest
neighbors in liquid depends on density and temperature, we believe
that using the fixed number of nearest neighbors gives
qualitatively correct results.

So, our point of view is that the definition of the structural
anomaly via $S_{\rm ex}$ is the most adequate at least for two
reasons: (i) only this definition has the well-defined
thermodynamic meaning, and (ii) there is a strict thermodynamic
relation between the density anomaly and the excess entropy
anomaly.

\section{IV. Conclusions}

The present paper discusses different definitions of the
structural anomaly in fluids, which one can find in literature, on
the basis of the core-softened potential, introduced in our
previous papers
\cite{silicalike,jcpsequence,specialtopics,we1,we2,we4,we3,RCR}.
We calculate the regions of structural anomalies with the use of
different definitions and compare them with each other. We show
that depending on the definition these regions can look rather
different. In our opinion, the most consistent definition of the
the structural anomaly in fluids is the one, based on the behavior
of the excess entropy, because only this definition has the
well-defined thermodynamic meaning. The other definitions can be
also used depending on the physical sense of the problem and
convenience of calculations; however, it is always necessary to
remember that different definitions can lead to drastically
different results.

\bigskip

\begin{acknowledgments}
Yu.F. and E.T. thank the Russian Scientific Center Kurchatov
Institute and Joint Supercomputing Center of Russian Academy of
Science for computational facilities. The work was supported in
part by the Russian Foundation for Basic Research (Grants No
14-02-00451, 13-02-91177, 13-02-12008, 13-02-00579, 13-02-00913 and 13-02-01099)
and the Ministry of  Education and Science of Russian Federation (project MK-2099.2013.2).
\end{acknowledgments}

\end{document}